\documentclass[a4paper,11pt]{article}
\usepackage{jinstpub} 
\usepackage{lineno}

\title{\boldmath Overview of the HL-LHC CMS outer tracker upgrade
and lessons learned from 2S module production}

\author{F. Zhang for the Tracker Group of the CMS Collaboration\footnote{Copyright 2026 CERN for the benefit of the CMS Collaboration. Reproduction of this article or parts of it is allowed as specified in the CC-BY-4.0 license.}}

\affiliation{IIHE, Universit\'e Libre de Bruxelles\\
Brussels, Belgium}

\emailAdd{fengwang@cern.ch}

\abstract{For operation at the High-Luminosity LHC, the CMS collaboration is developing a significantly enhanced silicon detector system with improved radiation hardness, higher granularity, and extended coverage up to a pseudorapidity of $|\eta|\approx$ 4. The Outer Tracker will be equipped with two types of modules (2S and PS) that will provide tracking information to the L1 trigger of the HL-LHC CMS detector, by rejecting signals from particles below a certain transverse momentum threshold. Hence, these modules must be assembled with high precision and have to pass strict requirements. In this note, an overview of 2S module production is presented, as well as representative examples of problems and experience acquired during module assembly and performance tests.}


\arxivnumber{2609.38026} 

\begin{document}
\maketitle
\flushbottom

\section{Introduction}
\label{sec:intro}
\begin{figure}[htbp]
\centering
\includegraphics[width=.98\textwidth]{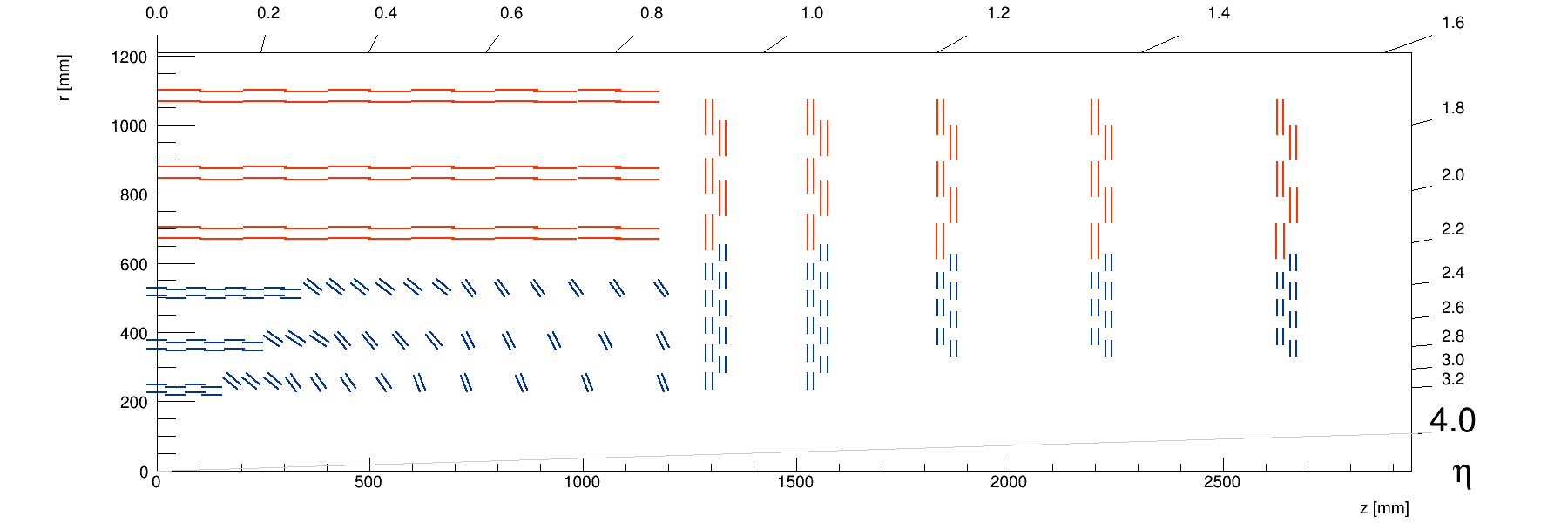}
\caption{Sketch of one quarter of the CMS Phase-2 Outer Tracker layout in r-z view. The blue and red lines represent the two types of modules, PS modules equipped with one macro-pixel sensor and one strip sensor and 2S modules equipped with two strip sensors, respectively. The figure is from~\cite{c}. \label{fig:otlayout}}
\end{figure}

The High-Luminosity Large Hadron Collider (HL-LHC)~\cite{a} is expected to start operation from 2030 and to deliver instantaneous luminosities of up to 7.5 $\times$ 10$^{34}$ $\textup{cm}^{2}\textup{s}^{-1}$. In order to cope with high pileup conditions of about 140 to 200 collisions per bunch crossing, the tracker of the CMS experiment~\cite{b} will be replaced with a new one~\cite{c}. The new system is designed to be radiation tolerant up to an integrated luminosity of 3500 fb$^{-1}$ expected after ten years of HL-LHC operation and to provide enhanced granularity for efficient particle tracking in high pileup conditions. Compared to the legacy CMS tracker, it has lower material budget and increased coverage in pseudorapidity. It consists of a silicon pixel detector (Inner Tracker) and a combined silicon pixel and strip detector (Outer Tracker or OT). The Inner Tracker will not be discussed further in this contribution. For the first time at a hadron collider, information from the OT will contribute to the CMS Level-1 (L1) trigger decision~\cite{d}. The layout of the OT is shown in Fig.~\ref{fig:otlayout}.

\section{Overview of modules for the CMS Phase-2 Outer Tracker}
\begin{figure}[htbp]
\centering
\includegraphics[width=.6\textwidth]{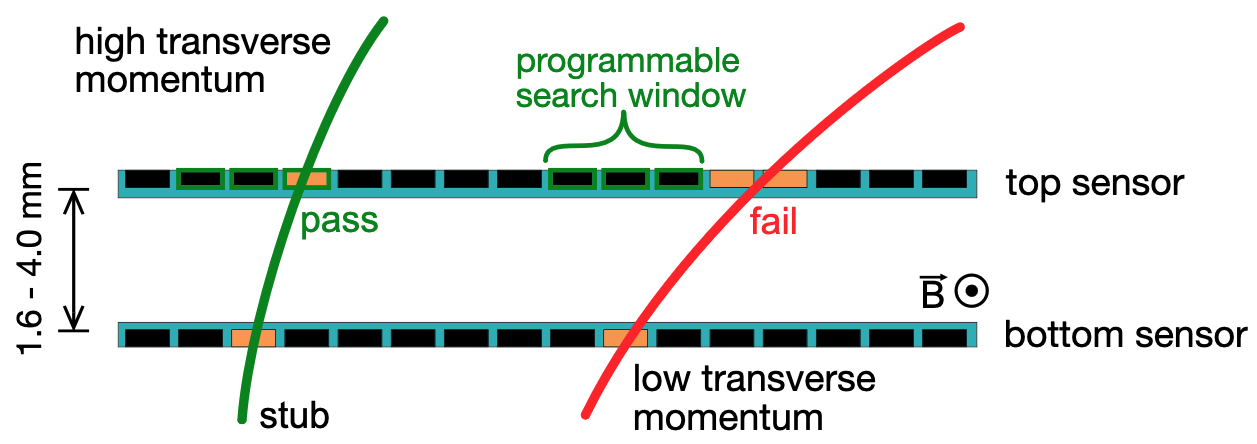}
\caption{Principle of the p$_{\textup T}$ discrimination with two parallel silicon strip sensors in a magnetic field~\cite{c}. Silicon strips oriented into the viewing plane with a pitch are represented by the black boxes. The detected clusters are shown in apricot boxes. The programmable search windows are in green contour. The left green curve shows a high p$_{\textup T}$ traversing charged particle and the right red curve stands for a low p$_{\textup T}$ particle. \label{fig:ptmodule}}
\end{figure}

\begin{figure}[htbp]
\centering
\includegraphics[width=.46\textwidth]{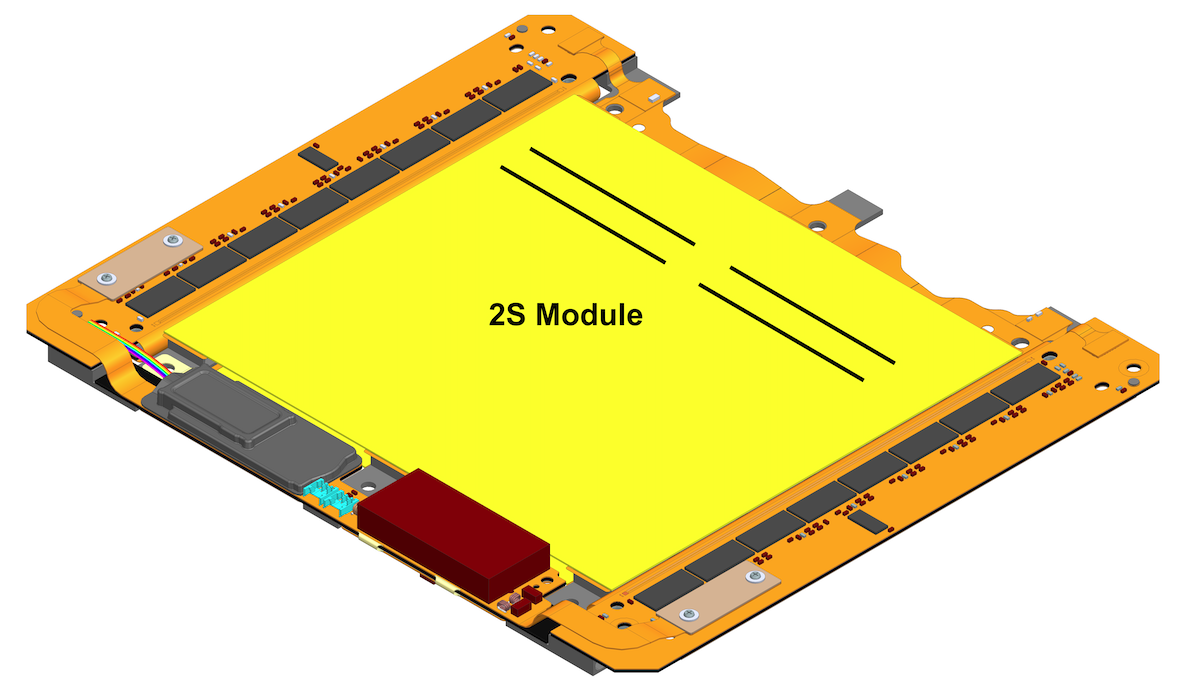}
\includegraphics[width=.46\textwidth]{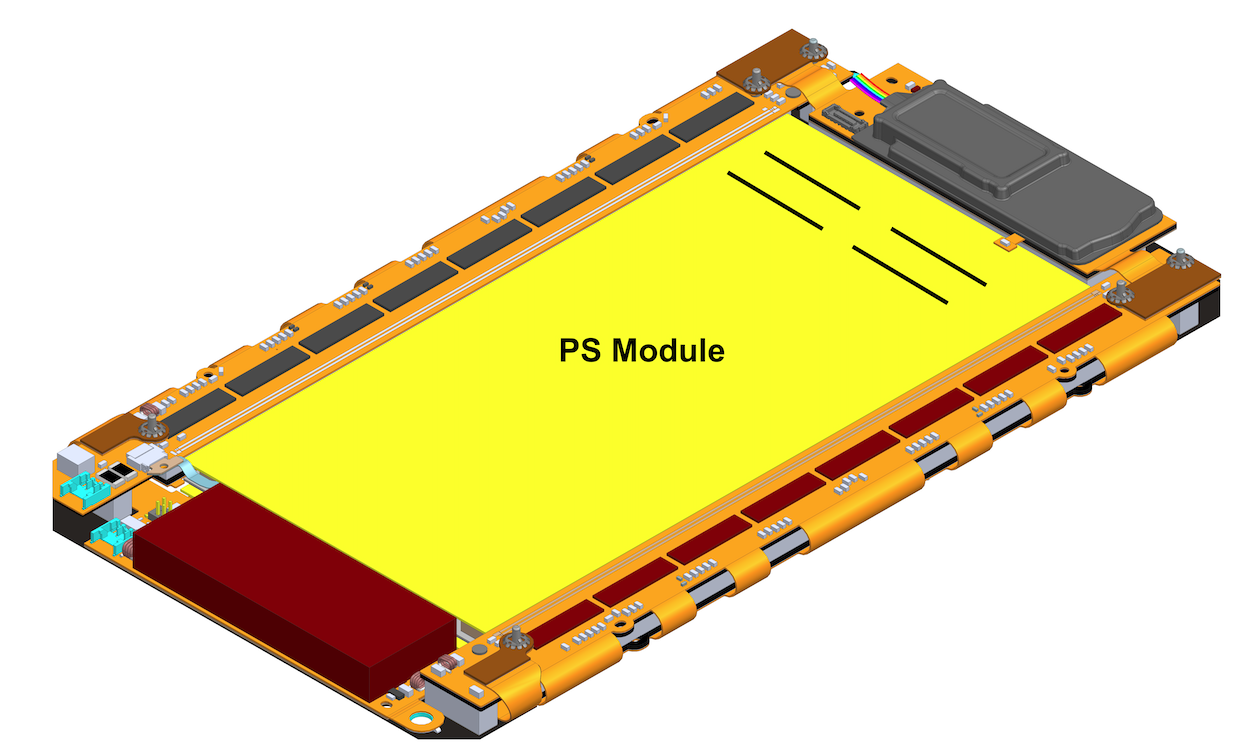}
\caption{The 3D view of a 2S module (left) and PS module (right). The black lines indicate the directions of the silicon strips. The figure was updated from~\cite{c}. \label{fig:ptmodules}}
\end{figure}

The OT will be equipped with two types of modules to facilitate the delivery of tracks for the L1 trigger. The modules are designed to cope with the bandwidth constraints by implementing an on-module discrimination on the transverse momentum (p$_{\textup T}$) of each particle. The presence of particles with higher p$_{\textup T}$ is an indication of a hard collision indicating that a potentially interesting physics event may have occurred. Figure~\ref{fig:ptmodule} shows a sketch of the stub mechanism for a p$_{\textup T}$ module: When a cluster is detected in the bottom sensor (seed layer), the closest cluster within a programmable search window on the top sensor (correlation layer) generates a cluster pair. Stubs are formed from
cluster pairs compatible with particle tracks above the chosen p$_{\textup T}$ threshold. By applying a p$_{\textup T}$ threshold of 2 GeV, the amount of data can be reduced by about 90\%.  Figure~\ref{fig:ptmodules} shows the final version of the 2S modules and PS modules. The design of these two types of modules were refined based on detailed studies of different generations of prototypes.

\clearpage
\section{Partition overview of CMS Phase-2 Outer Tracker}
\begin{figure}[htbp]
\centering
\includegraphics[width=.42\textwidth]{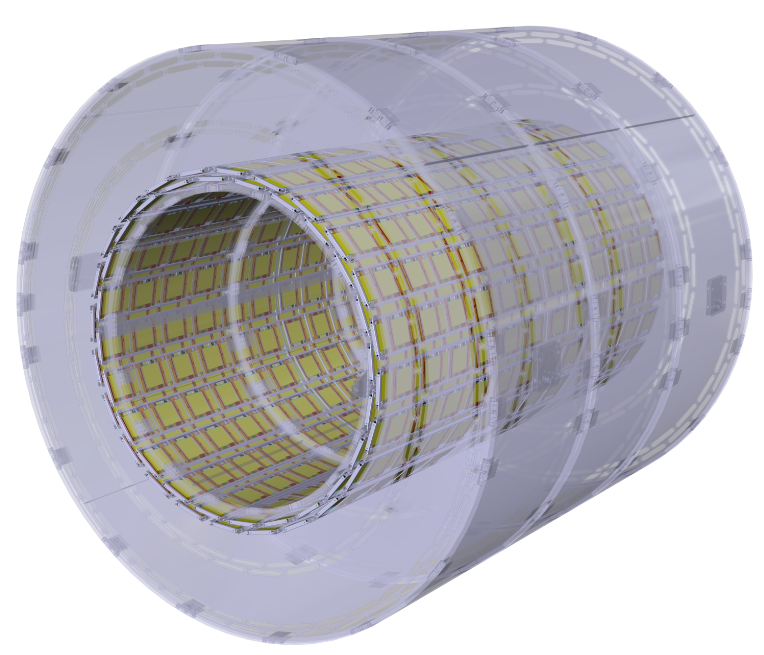}
\includegraphics[width=.37\textwidth]{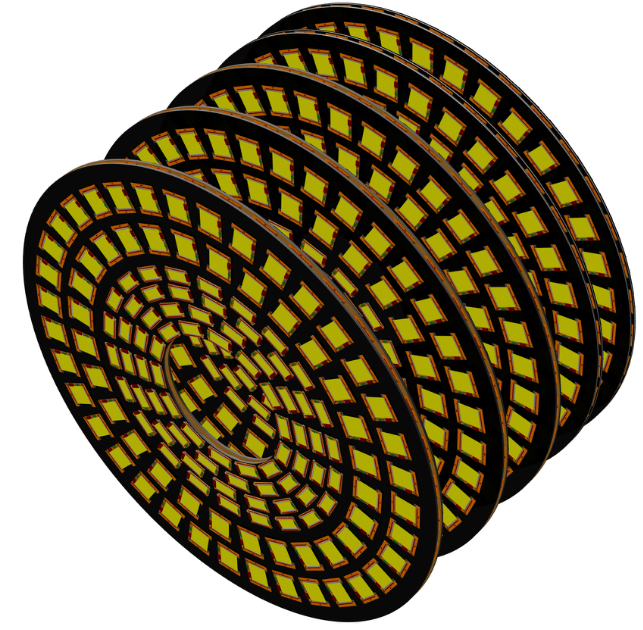}
\caption{Left: The ladders of the innermost layer of TB2S as installed in the support wheel; Right: A TEDD unit consisting of five double-discs. Each double-disc consists of four dees. The figure is from~\cite{c}. \label{fig:tb2s_tedd}}
\end{figure}

The OT covers up to $|\eta|\approx$ 2.4, and is made of 3 partitions: TB2S, TBPS, TEDD. The TB2S will be equipped with 4416 2S modules. A total of 2872 PS modules will be mounted in the TBPS. The TEDD will contain a mixture of PS and 2S modules for a total of 5912 modules. Figure~\ref{fig:tb2s_tedd} shows a part of TB2S and TEDD.

\section{2S module production and qualification test}
\begin{figure}[htbp]
\centering
\includegraphics[width=.85\textwidth]{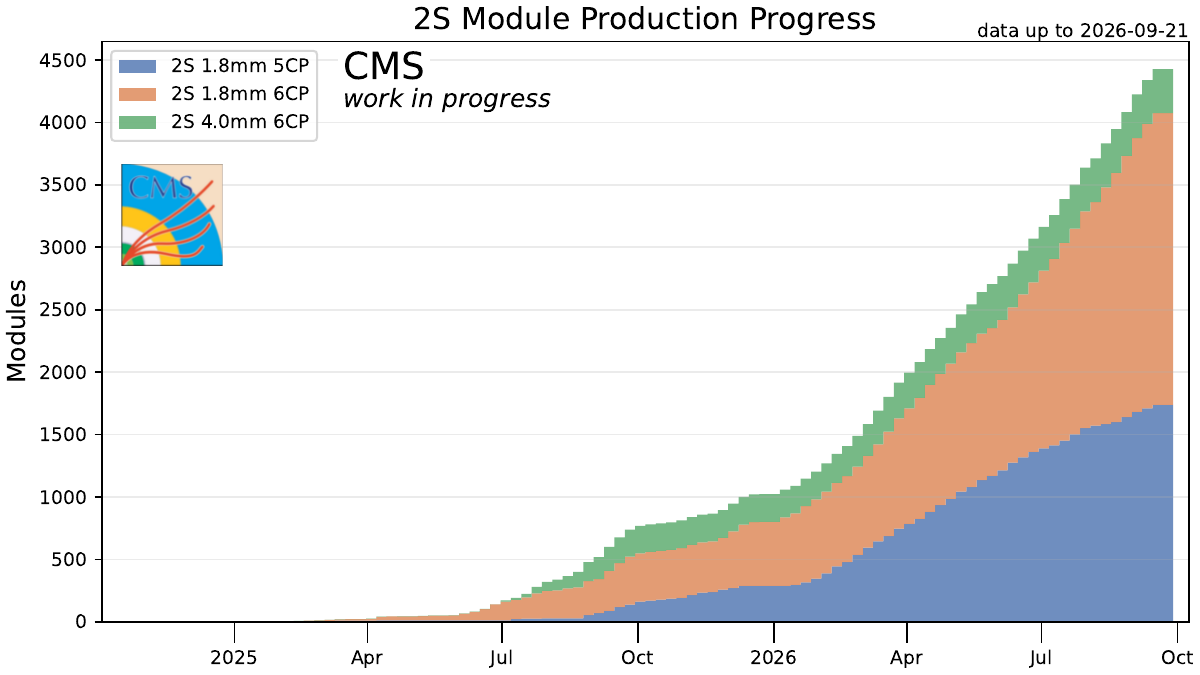}
\caption{The history of global 2S module production progress by variant.\label{fig:2s_assemble_progress}}
\end{figure}

The CMS Tracker Group has been coordinating 12 production centers across Asia, Europe, and North America to produce hundreds of modules per week. The 2S module production progress is shown in Fig.~\ref{fig:2s_assemble_progress}.

\begin{figure}[htbp]
\centering
\includegraphics[width=.55\textwidth]{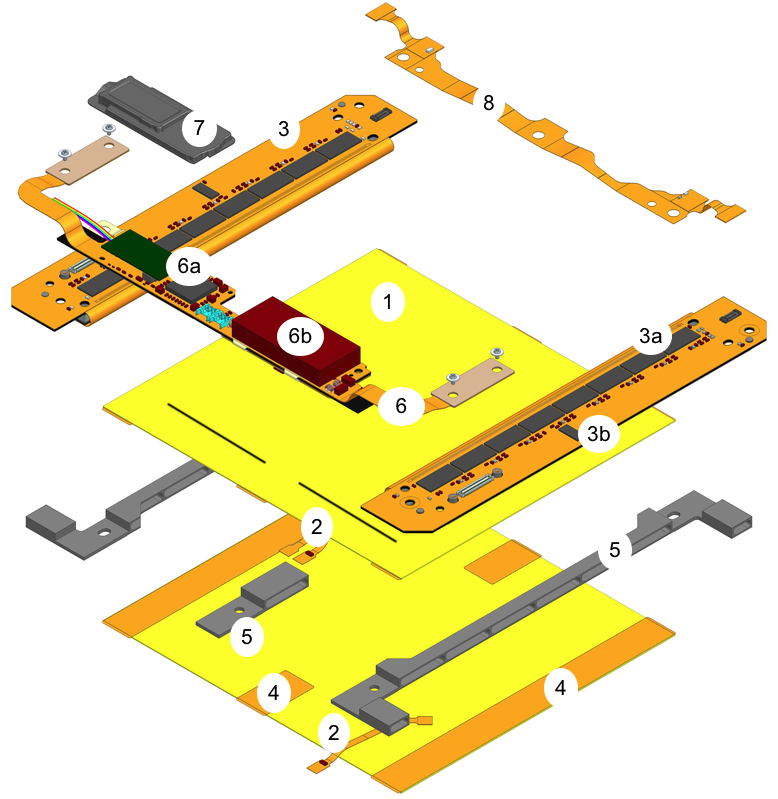}
\caption{An exploded view of a 2S module (4 mm 6 CP variant). It consists of two identical silicon strip sensors (top and bottom), with lines representing the strips for illustration (1), pigtails (2) bonded at the back side of each sensor for applying bias voltage, two Front-End Hybrids FEH (3),
each with eight CMS Binary Chips CBC (3a) and a Concentrator Integrated Circuit for data compression (3b), and Kapton-isolated (4) carbon fiber reinforced aluminum bridges (5) separating the two sensors. On the front side, there is a SErvice Hybrid SEH (6) with a Versatile Link Plus Transceiver, a low-power Gigabit Transceiver (6a), and low-voltage DC-DC converters below a shield (6b), and a light shield (7). A ground balancer (8) is placed on the other side of the module. The figure was updated from~\cite{c}.\label{fig:2s_explode_view}}
\end{figure}

\begin{figure}[htbp]
\centering
\includegraphics[width=.95\textwidth]{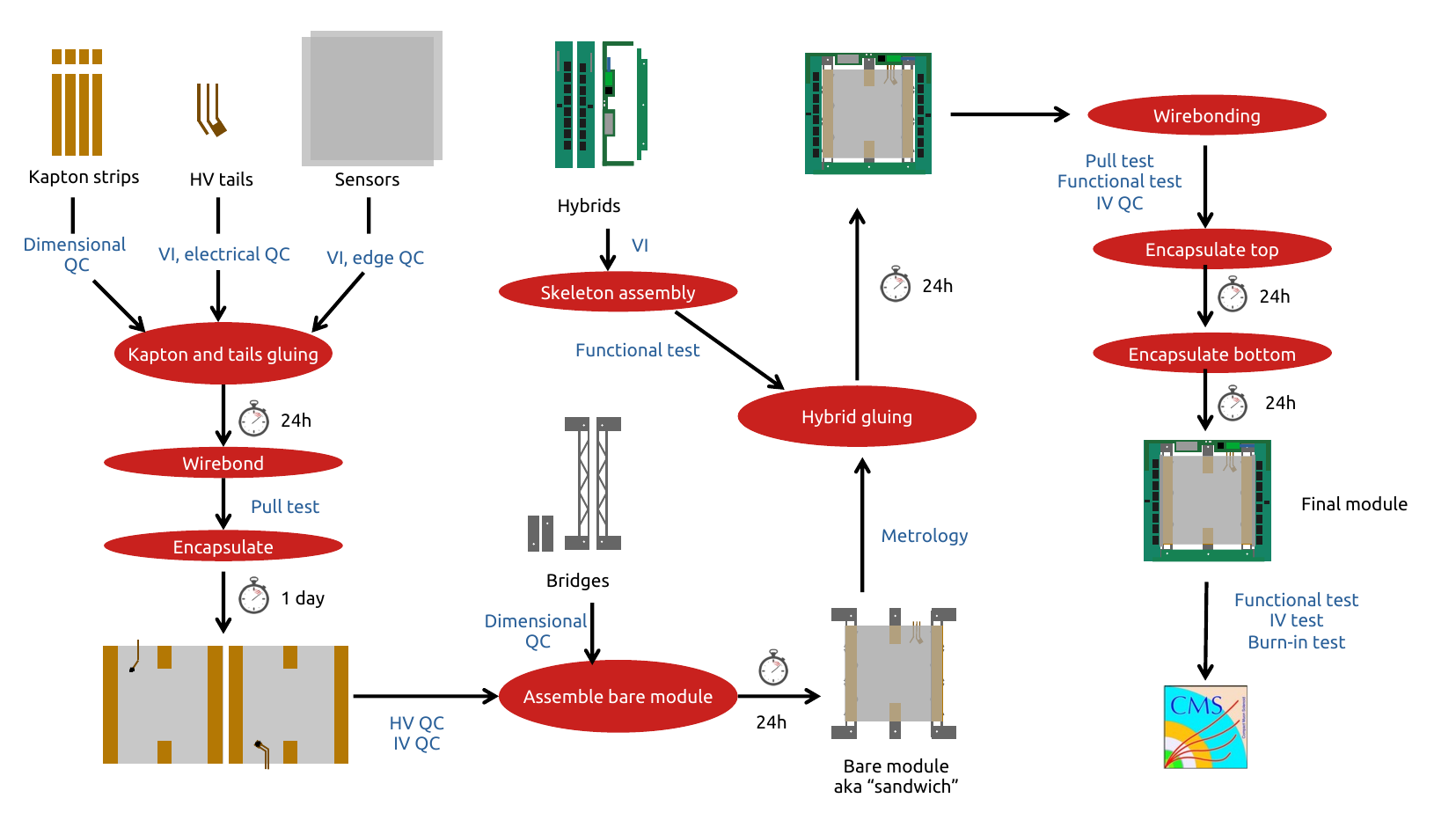}
\caption{Sketch of 2S module assembly and qualification test procedure during the module production.\label{fig:2s_assemble_qc}}
\end{figure}

\begin{figure}[htbp]
\centering
\includegraphics[width=.9\textwidth]{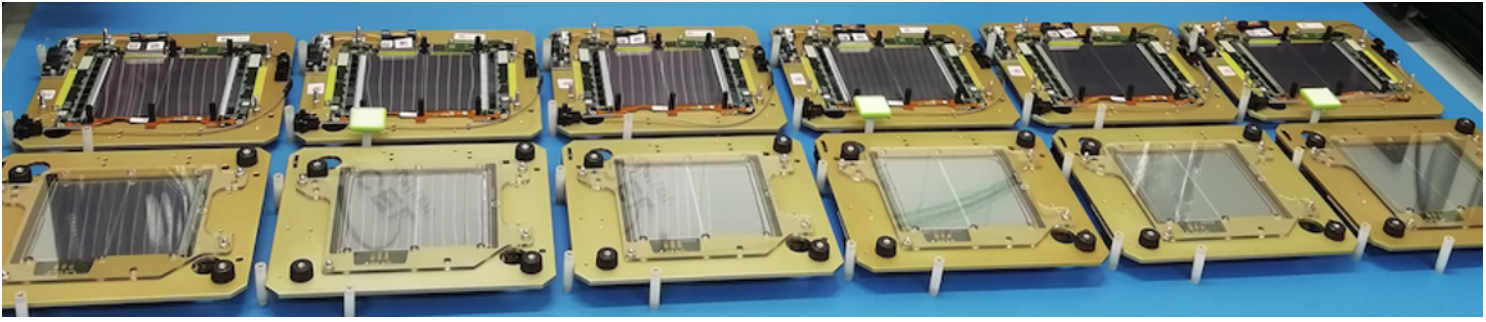}
\caption{A photo of 12 2S modules with their bond-wires encapsulated at the Brussels production center. \label{fig:2s_photo}}
\end{figure}

During the production, 3 different variants of 2S modules are being built. There are two different sensor spacings: 1.8 mm and 4 mm, chosen for efficient low-momentum track rejection at different locations in the detector. Furthermore, modules can come with 5 or 6 cooling points (CP) to guarantee sufficient cooling at any location. The three module variants built are: 1.8 mm 5 CP, 1.8 mm 6 CP, and 4 mm 6 CP. The 1.8 mm variants will be placed in the barrel and outer radius endcap regions, while the 4 mm variants will be installed in the inner radius endcap area. Figure~\ref{fig:2s_explode_view} shows an exploded view of a 4 mm 6 CP module. In order to guarantee the performance of the new CMS tracker, each module must be built in a strictly quality-controlled pipeline. Figure~\ref{fig:2s_assemble_qc} shows the standardized 2S module assembly and qualification tests during mass production. As an example, 12 modules with their bond-wires encapsulated are shown in Fig.~\ref{fig:2s_photo}.

\begin{figure}[htbp]
\centering
\includegraphics[width=.9\textwidth]{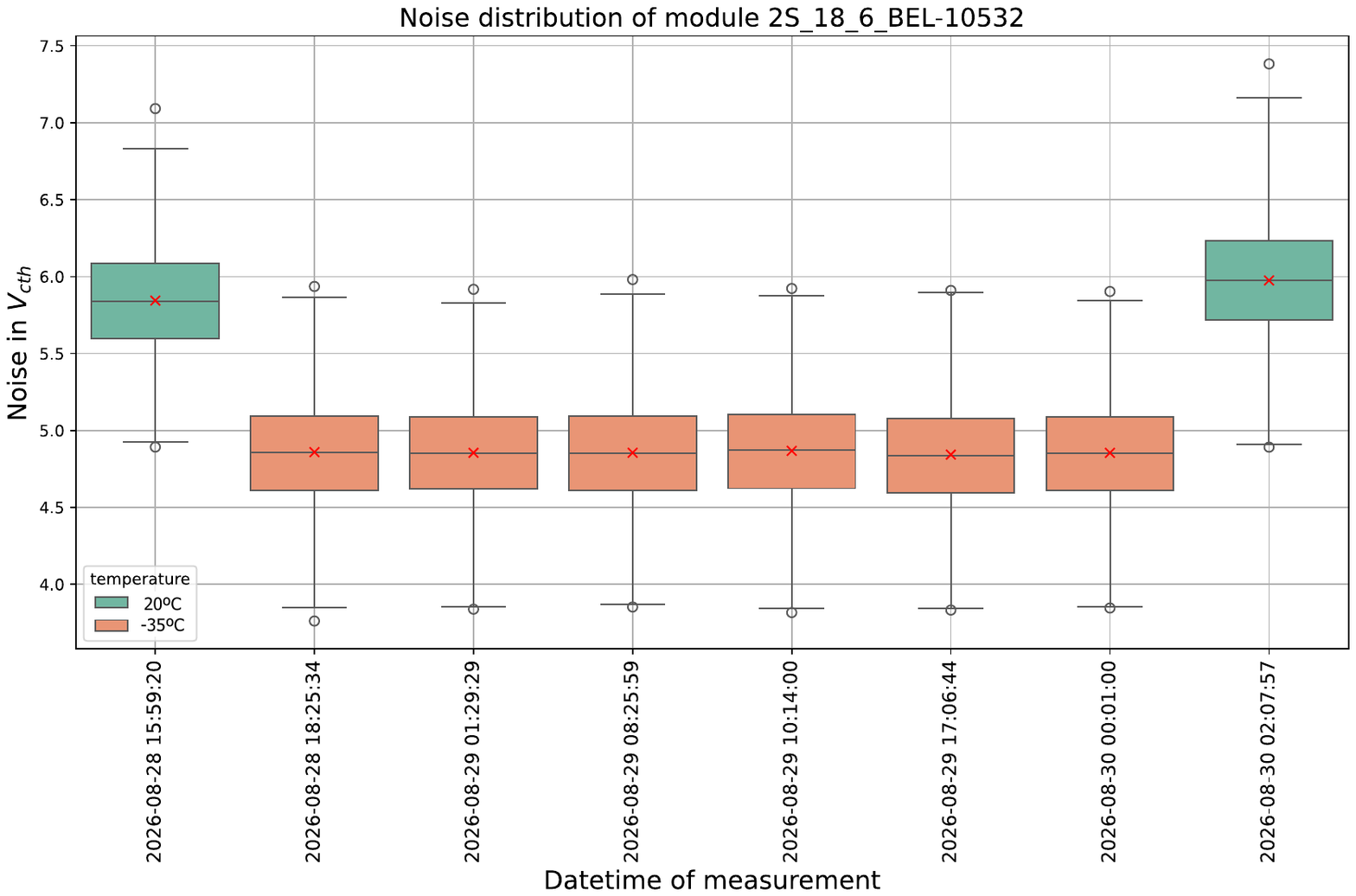}
\caption{Average and medium noise level of an assembled 2S module, measured twice at 20 $^{\circ}$C and 6 times at -35 $^{\circ}$C in the Belgian production center. The box contains the noise of 50\% of the strips. The line inside each box represents the median. The red cross is the mean. The whisker contains [0.02, 99.98] \% of all the strips. The lowest and highest strips are shown outside as circles. \label{fig:2s_module_noise_burnin}}
\end{figure}

\newpage
The 2S modules will be operated at a temperature of -35 $^{\circ}$C in the HL-LHC CMS detector. During the production, each module is required to undergo 16--18 thermal cycles between 20 $^{\circ}$C and -35 $^{\circ}$C during the so-called Burn-in test. An electronic readout test is conducted at both temperatures, in order to examine the mechanical robustness and the stability of the electronic performance of each module. In general, the electronic noise in cold is lower than in warm. An example of the measured average electronic noise of a good 2S module from a Burn-in test is shown in Fig.~\ref{fig:2s_module_noise_burnin}. The electronic noise distribution was measured twice in 20 $^{\circ}$C and 6 times in -35 $^{\circ}$C during 16 thermal cycles. 
For each of the tests described above, the output is analyzed and assigned a grade using a central application called Phase-2 Outer Tracker Analyzer of Test Outputs (POTATO)~\cite{e}. This is a quality-assurance and quality-control software tool developed using C++ to aid the coordination of the Phase-2 OT module production and evaluation process.

\section{Lessons learned from 2S module production}
During the production period, problems can occur during each assembly step, which could affect the final performance of the assembled module. It is crucial to understand the cause of these and address them as early as possible. Although it is unavoidable that human handling mistakes occur, one needs to minimize them as much as possible. In the following paragraphs we will discuss three representative lessons learned during the 2S module production campaign.

\subsection{Lesson 1: Scratches}
During the handling procedures in the production pipeline, such as IV measurements, making a bare module (two sensors are glued to Aluminum-Carbon Fiber bridges), gluing a bare module with hybrids, scratches could be formed by hard objects like needle-shape probes and jig pins. If a scratch is deep enough it can create a pin hole leading to an early breakdown of the module under bias as well as noisy strips. When a scratch across strips is shallow, the scratched strips can still show normal noise levels during the electric readout tests. As long as the leakage current of a scratched module is below 12 $\mu$A when it is biased at 800 V, the module is still acceptable for the final detector. For a module with scratched sensors showing an IV curve out of specification, removing bond-wires of the noisy strips has been attempted, which will reduce the leakage current flowing from strips to the amplifiers, and improve the IV characteristics of the sensor significantly. 

\subsection{Lesson 2: Electric sparks}
When a 2S module is biased at a high voltage level such as 800 V, it is crucial to control the electric isolation between the edge of the sensor at bias potential and the auxiliary electronics such as the FEH and SEH at ground potential. During the production, electric sparks between the sensor and the edge of the carbon-fiber stiffener have been observed, which led to high current flowing into the CBC chips and damaged them. Due to a manufacturing defect, the edge cross-section of an SEH can contain protruding carbon fibers that might be overlooked by the manufacturers. These are difficult to remove during module production. When such fibers are present, the air gap of about 400 $\mu$m is not capable to hold an electric potential of 800 V. This issue has been addressed by attaching a layer of Kapton isolator between the sensor and the SEH of each module during the production.

\subsection{Lesson 3: Unseen conductive contamination}
Conductive contamination happens during the handling procedure when the module carriers or the assembly jigs are not clean enough. Depending on the extent of contamination, an electrical test of an assembled module can still give a good result or otherwise show an IV breakdown. The symptom is that the HV current of one sensor starts rising sharply when the bias voltage is increasing to a certain level below 800 V. To treat it, one first retests it and verifies that the IV breakdown is consistent. After identifying the sensor in which the issue occurs, the module is placed under an ion blower for 10--15 minutes. If this does not help, the module is biased towards 800 V with a maximum HV current threshold (50 $\mu$A) in a very low relative humidity environment ($<$1\%) for 30--60 minutes. Some module IV curves return to normal after this. If not, Electrostatic Discharge ESD-safe tissues with ethanol are used to carefully wipe the edges of the problematic sensor. These measures are effective to treat modules with electric conductive contamination that leads to an early IV breakdown.

\section{Module production database}
During the module mass production, a large amount of data is generated. It is crucial to keep track of lots of information relating to logistics, assembly logs, measurement records, environmental data, remaining issues and so on. Besides the global management database, a local database is necessary for a production center. For example, at the Brussels production center, a modular local database, called Global Assembly Bookkeeping with Relational Image and Event Logging (GABRIEL) was developed. It not only stores all the production data, but also correlates them to help coordinate tasks. For example, if some parts are tested out of specification, GABRIEL will notify the operator team to avoid using them in the pipeline. In practice, the daily production data are written by operators into GABRIEL, which automatically uploads those data to the global database overnight.

\section{Summary}
This note gave an overview of the Phase-2 CMS Outer Tracker of the HL-LHC. It introduced the concept of two different types of p$_{\textup T}$ modules, and different sections of the detector where the modules will be installed. Besides showing the progress of the ongoing 2S module production campaign, this note also illustrated the procedure of 2S module assembly and qualification tests. Three representative lessons learned from the production, including scratches, electric sparks, unseen conductive contamination, were discussed. In addition, a modular database developed by the Belgian production center team was presented. The mass production of 2S modules is expected to be completed in 2027.

\end{document}